\documentstyle[11pt,aaspp4,epsfig]{article}

\begin{document}

\title
{\bf STELLAR POPULATIONS AT LARGE REDSHIFTS\footnote{To be published
in the 1999 revised edition of 
{\it Stellar Populations (IAU Symposium No.\ 164)}, eds.~P. van der Kruit
and G. Gilmore (Dordrecht: Kluwer).} }

\author{Robert W. O'Connell}
\affil{Astronomy Department \\
           University of Virginia\\ 
           Charlottesville, VA 22903-0818   USA}


\def\alwaysmath#1{\ifmmode {#1}
                  \else {$#1\mkern-5mu$} \fi} 
\catcode`\@=11
\def\gsim{\alwaysmath{\mathrel{\mathpalette\@versim>}}}
\def\lsim{\alwaysmath{\mathrel{\mathpalette\@versim<}}}
\def\@versim#1#2{\lower 2.9truept \vbox{\baselineskip 0pt \lineskip
    0.5truept \ialign{$\m@th#1\hfil##\hfil$\crcr#2\crcr\sim\crcr}}}
    \catcode`\@=12
    
\def\arcsec{\ifmmode {'' }\else $'' $\fi} 
\def\arcmin{\ifmmode {' }\else $' $\fi} 


\baselineskip=11pt

\begin{abstract} 
Rapid progress is now being made in the study of stellar populations
of galaxies at large lookback times, both in dense clusters and the
field.  Dramatic transformations in star formation histories (even
morphologies) appear to prevail among all types of galaxies and in all
environments, and these can be traced to relatively recent times ($z
\sim 0.2$).  This article updates to December 1998 a review made in
1994 to cover the recent watershed of observational results emerging
from the Hubble Deep Field and deep surveys made with large
ground-based telscopes.  
\end{abstract}

\section{Introduction}

With one exception, evidence for evolution in nature is
indirect and circum\-stan\-tial in that it is not based on
contemporaneous observations.  The exception, remarkably enough, is
evidence for evolution of the universe itself.  Courtesy of the vast
scale of the universe compared to the finite speed of light,
astronomers are able to see directly what was happening billions of
years ago.  However, the distant universe is faint enough that we have
only recently been able to exploit this opportunity.  In fact, at the
time of the Vatican Conference in 1957, as Adriaan Blaauw noted in his
historical review, ``Nothing was said about galaxy evolution because
nothing sensible could be said.''  Today, one can sensibly conclude
that galaxy evolution has been detected, which is a result of enormous
importance, but the number of other sensible remarks which can be made
is perhaps limited.

``Galaxy evolution'' is not a very well defined term.  Workers
generally do not use it to mean the evolution of individual galaxies,
which has been recognized an inevitable since the beginning of galaxy
research (Hubble 1936).  Instead, they mean a systematic change with
cosmic time in the statistical distribution of galaxies over their
global characteristics.  In principle, evolution can be demonstrated
by a change in any one property of the galaxy sample.  However, much
more information is required to interpret the character of
that evolution, particularly if it involves stellar populations.  

This statistical definition requires that we compare local ($z\,
\lsim\, 0.05$, say) samples of galaxies to distant samples, both
chosen to reflect representative conditions.  If we are not careful in
matching the two samples, we can readily misinterpret the evidence for
evolution.  Further, the physical nature of evolution can only be
understood by comparing the two epochs in detail.  Thus, a
comprehensive assessment of our local environment is a key ingredient
in evaluating the evolution of distant galaxies.  This assessment is
currently rather limited, and I will comment briefly on that below.

Rapidly improved technology has allowed the study of galaxies at large
lookback times to blossom.  Nonetheless, the distance moduli involved
($\sim 45$ at $z = 1$) challenge even the best instrumentation.
Consequently the {\it dimensionality\/} of stellar population analysis
of distant objects has been limited mainly to the star formation
history (SFH), with a few recent studies adding morphological
information.  Knowledge of abundances (other than interstellar gas),
kinematics, or IMF's is rudimentary.

The seminal result in the field was the discovery by Butcher
\& Oemler (1978) that galaxy clusters at modest redshifts of $z \sim
0.4$ contained more blue objects than local clusters.  Doubts about
field contamination and interpretation of the colors were resolved by
spectroscopy (e.g. Dressler \& Gunn 1982) and larger samples, and the
``Butcher-Oemler Effect'' is now regarded as unambiguous evidence for
the evolution of galaxies in rich cluster environments over the past
$\sim 5$ Gyr.  Simultaneously, studies of field galaxies (e.g. Tyson
\& Jarvis 1979) revealed a rapid
increase in number to $m \sim 24$ and a shift to bluer colors, which
most workers regard as good evidence for evolution in the field.  

In this review, I will concentrate on observations of rich clusters,
which provide the best fiducial environment available.  There are many
interesting aspects of high redshift populations to which I cannot do
justice, including very distant radio galaxies, gravitational
lensing, and the magnitude-redshift relation.

\section{Evolution of Disk Galaxies}

\subsection{Populations In Local Disk Samples  }

\noindent Disk galaxies (i.e. Hubble types later than S0) constitute
60--70\% of field and group samples; only in clusters does their
proportion drop below 50\% (Oemler 1992).  However, despite their
abundance (and the fact that we live in one), we have a relatively
poor understanding of their evolution, at least on the global scale
relevant for comparison with distant galaxies.  There are both
fundamental and technical reasons for this.  First, these are
multicomponent systems, and their integrated properties are influenced
by two very different populations: the disk and the bulge.  Even
within disks, there can be strong gradients in SFH and metallicities.  

On the technical side, we have probably been misled by the ease with
which we can deduce the {\it current\/} massive star formation rate
(e.g. from emission lines).  This applies only to ionizing
populations, which have ages $\lsim\, 5$ Myr, or 0.05\% of a galaxy's
lifetime.  The other 99.95\% is a tougher proposition, and the most
readily available input data---broad-band colors---are inadequate to
the task.  Broad-band colors can constrain the gross SFH
(e.g.~Gallagher et al.~1984), but it is now obvious that they contain
far too little information to be useful in resolving the details of
interest (e.g.~smooth star formation vs.~bursts vs.~quenching) in the
context of the early universe.  Although the colors of most disks are
consistent with smooth, exponentially declining SFH's in very old
systems (e.g.~Larson \&
Tinsley 1978), they are also consistent with a variety of radically
different histories (Tinsley 1980, Schweizer \& Seitzer 1992,
Fritze-v.~Alvensleben \& Gerhard 1993).  

Since broad-band measures
seriously under-constrain stellar populations, analyses using them must
provide for automatic exploration of the full alternative solution
space.  One method was described by Wu et al.~(1980), but such
algorithms have been little used.  A good example of the controversies
which can arise in their absence is the debate over the age of distant
radio galaxies such as 0902$+$32 (Lilly 1988, Chambers
\& Charlot 1990).  

There are few published studies on the general ability of integrated
light measures to infer the SFH of multi-component populations.
One useful benchmark is the work of Pickles \& van der Kruit
(1990), who found that S/N $> 20$ in 13.5 \AA\ bins between 3600 and
10000 \AA\ was needed to distinguish three population components,
each specified by their age and metallicity.  That is, a problem
involving 6 unknowns required over 70 times as many data points.  When
strong redundancies are eliminated, the data requirements can probably
be significantly reduced.  However, the fact that stellar energy
distributions are not mathematically independent (cf.~Silva 1991),
which is the root cause of the problem here, implies that high
precision, multi-band spectrometry is essential for analysis of
composite populations.  Only recently has such data become available
with large apertures on a large sample of disk galaxies (Kennicutt
1992), and their implications for the SFH have yet to be
systematically explored.

\subsection{Distant Field \& Cluster Disk Systems}

The smooth, exponential SFH's which were favored for spirals ten years
ago predict little change in spectral properties with redshift until
one reaches almost the formation epoch (e.g.~Tinsley 1980).  Field
counts seem consistent with this sort of ``mild'' evolution
(Bruzual \& Kron 1980, Tyson 1988), with no evidence for a unique
formation epoch at $z\, \lsim\, 5$.  But spectra for $m\,
\lsim\, 23$ (e.g.~Lilly et al.~1991, Broadhurst et al.~1992) yield a
surprisingly small $\bar z \sim 0.4$ and few high redshift
($z\,\gsim\,0.75$) objects.  This implies a major change in field
properties by $z \sim 0.5$ (a 2--5$\times$ increase in the number of
blue objects with L$<$L$^*$); the spectra also suggest an enhancement
of starburst activity.  The best explanation is non-conservation of
galaxy numbers, probably indicating widespread merging since $z \sim
1$ of smaller objects into larger ones with accompanying starbursts.
This is controversial, however, and the ``conservative'' school argues
that the observations can be explained without evolution of any kind,
within the uncertainties of local samples and population modeling (Koo
\& Kron 1992, Koo et al.~1993).  My impression is that the
no-evolution models do not fit the field data well despite rather
extreme input assumptions.  However, this is the subject of other
reviews in this volume (Ellis and Gardner), so I will not discuss it
further.  

In the clusters, the changes found at modest redshifts by Butcher \&
Oemler (1978), namely an increase in the fraction of galaxies which
are $\gsim 0.2$ mags bluer in (B--V) than typical E/S0 objects, were
also unanticipated in the ``mild'' evolutionary models.  In nearby
clusters, the blue fraction is only $\sim 2$\%, whereas at $z \sim
0.5$ it is $\sim 30$\%.  The distant clusters are actually similar to
the nearby field, and we now recognize that it is an
environmentally-driven {\it loss\/} of this blue population in nearby
dense clusters which defines the Butcher-Oemler Effect (Oemler 1992,
Dressler et al.~1994).  

The broad-band color range of the blue systems is comparable to that
of local, normal spirals.  It was therefore initially presumed that
they are normal disk galaxies.  Only recently has high resolution
imaging ($\lsim 0.5\arcsec$) with CFHT and HST confirmed that they are
indeed predominantly disks (Lavery et al.~1992, Couch et
al.~1994, Dressler et al.~1994).  However, spectroscopy (reviewed by
Dressler \& Gunn 1990) reveals that some 80\%  are not
normal disk galaxies.  Many appear to have abnormally high star
formation rates (starbursts) or to be in a quenched phase $\sim$1--3 
Gyr following a burst, so that the main sequence includes A stars but
does not extend to more massive types.

\begin{figure}

\centerline{\epsfig{file=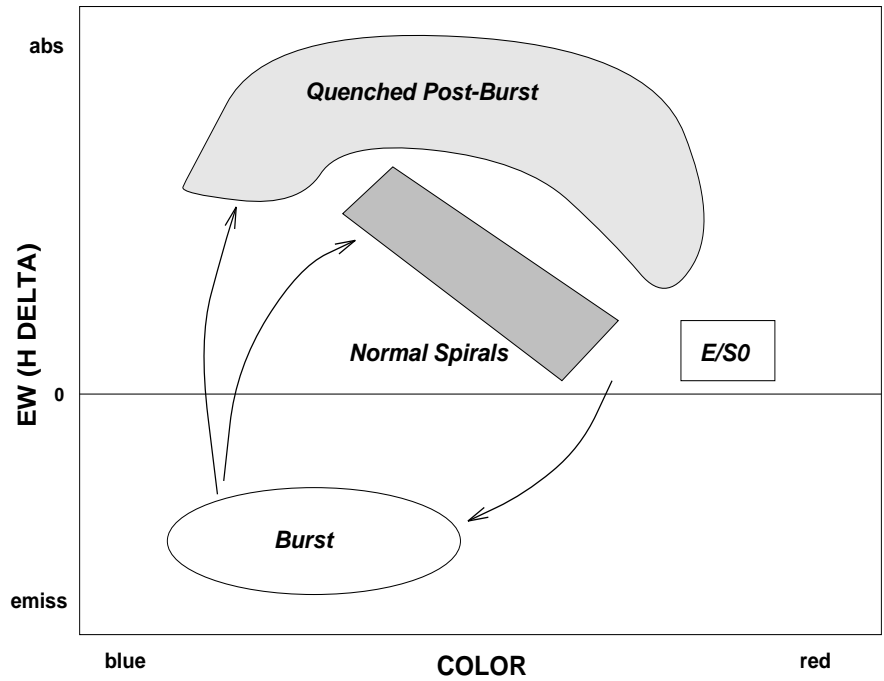,height=3.5in,width=4.6in}}

\vspace{4pt}
\caption{ The location of normal galaxies and burst/post-burst systems
in a Balmer line vs. color diagram.  The arrows show evolutionary
trajectories during and after a starburst.  ``Normal spirals'' have
smoothly declining star formation rates.  In ``quenched'' systems
normal star formation is suppressed after the burst.  }
\end{figure}

Couch \& Sharples (1987) illustrated a particularly useful diagnostic
of the blue Butcher-Oemler systems.  They plot the net
equivalent width of H$\delta$, measured with $\lsim 10$\AA~resolution,
against color (see Fig. 1).  In this diagram, the burst and quenched
systems are well separated from normal spirals and quiescent E/S0's.
Although H$\delta$ is used here, a number of spectral
features in the 3500--4500\AA\ region would yield similar results.
Broad-band colors, however, would not suffice.

The unusual histories implied by the spectroscopy combined with the
recent imaging, which shows that many blue galaxies appear disturbed
or are in interacting pairs, strongly implicate interactions or
mergers in the Butcher-Oemler Effect.  Not all peculiar objects have
close companions (Dressler et al.~1994), but this does not necessarily
exclude interactions, as shown by the local example of M81/M82.  The
spiral galaxy M81 has almost certainly induced the starburst now
observed in M82's core; yet neither appears seriously disturbed at low
resolution in the optical (except at H$\alpha$), and they are separated
by 39 kpc (or $7\arcsec$ at $z \sim 0.3$).  The outer parts of M82
have a quenched spectrum.  This is a good example of a lingering
starburst, as discussed by Schweizer in this volume.

Whether interactions or mechanisms such as IGM ram-pressure
(Oemler 1992, Dressler \& Gunn 1990) drive the Butcher-Oemler Effect,
the destruction of cold gas disks was well advanced
by $z \sim 0.5$ (lookback time $\sim 5$ Gyr) in rich cluster environments.
The final products will be gas-poor, quiescent spirals and S0's 
(abundant in nearby clusters) and, if mergers are common, ellipticals.
The radial distributions discussed by Couch \& Sharples
(1987) suggest that E/S0's are the most common descendents.  

\section{ Evolution of E/S0 Systems}

\subsection{Local Ellipticals}

Any attempt to understand the histories of elliptical galaxies is
bedeviled by a dichotomy of properties.  On one hand, there is
impressive evidence for {\it homogeneity\/} among ellipticals, for
instance the good correlations between luminosity and color (Bower et
al.~1992) and between velocity dispersion and the Mg$_2$ line index
(Bender et al.~1993).  The traditional emphasis has been on such
regularities, which appear to argue for a single history for all E's,
as described by Renzini and Franx at this conference.  On the other
hand, there is equally strong evidence, beginning with indications that
low-luminosity ellipticals in the Local Group contained intermediate
age populations (Aaronson \& Mould 1980, O'Connell 1980), for {\it
heterogeneity\/} and consequently for diverse histories.  Insofar as
integrated light is concerned, interpretations are once again strongly
influenced by whether one uses broad-band colors or higher resolution
spectra (e.g.~Rose 1994).  I will not give details here,
since they have been covered in recent reviews (Freedman 1994,
O'Connell 1994) and were the subject of papers at this conference by
Bender, Faber, Freedman, and Schweizer.

The local evidence for diversity demonstrates that star formation in
E/S0 galaxies can be, though is not always, prolonged (to
$t\,\lsim\,5$ Gyr ago) and episodic and is often driven by
environmental effects.  Indeed, that would be entirely consistent with
the high redshift data, as described above.  The dichotomy of
properties arises at least partly from an environmental effect: in
denser regions, E/S0 galaxies appear to have evolved faster (Lucey,
this conference).  This does not entirely explain the regularities, as
Renzini and Franx have properly emphasized.  Nonetheless, the evidence
is strong that E/S0 galaxies are not uniformly old systems which
evolve in isolation.

\subsection{Evolution of The Red Envelope}

The ``red envelope'' is one of the phenomena suggesting the homogeneity
of E/S0 galaxies.  It is an empirical upper bound on the colors of
galaxies (Sandage \& Visvanathan 1978, O'Connell 1988).  At low
redshift the envelope lies at (B--V)$\sim 1.0$, it is well defined in
rich clusters, and it is occupied preferentially by luminous E/S0
galaxies.  This immediately suggests a test of evolution and galaxy
ages: how does the envelope change with redshift? 

Properly speaking, this does not test the evolution of {\it
elliptical\/} galaxies since morphologies are mostly unknown at $z\,
\gsim\, 0.2$.  The red envelope is occupied by the {\it least active
subset\/} of galaxies irrespective of morphology.  In the absence of
abnormalities (e.g.~AGN's, dust, line emission, unusual IMF's, etc.),
red envelope systems have experienced the smallest amount of star
formation during the preceding $\sim 3$ Gyr.  It is incorrect to
assume, as is not uncommon, that all red envelope systems are E/S0
galaxies or that all E/S0 galaxies fall near the envelope. It is also
misleading to use ``E/S0'' to designate spectral or color
classifications of distant galaxies whose morphologies are not known.

The red envelope test on quiescent systems is important because it is
very difficult to place limits on the early SFH of galaxies
with recent star forming activity, such as those discussed in \S 2.
A schematic version of the test is shown in Fig.~2.
One uses a color or gross spectral feature such as the 4000 \AA\ Break,
preferentially measured in the rest frame of the distant galaxies to
avoid ``k-corrections''.  The test is empirical and differential, so
broad-band measures are suitable.  Current evolutionary synthesis models
(e.g.~Bruzual \& Charlot 1993, Worthey 1994, Bressan et
al.~1994) agree on the rate of change of such colors even if they
don't agree on  absolute values (Worthey 1994).  For the 
case shown, the models predict $\partial (U -J) / \partial
\log t \sim 1.4$ mag/dex.  Despite its simplicity, there are many
obstacles to applying the test (cf.~Arag\'on-Salamanca et al.~1993). 
One complication with likely astrophysical content is the
occurrence of ``red interlopers''---objects with colors too red for a
population younger than the age of the universe and therefore probably
reflecting an abnormality such as dust or an AGN.

\begin{figure}

\centerline{\epsfig{file=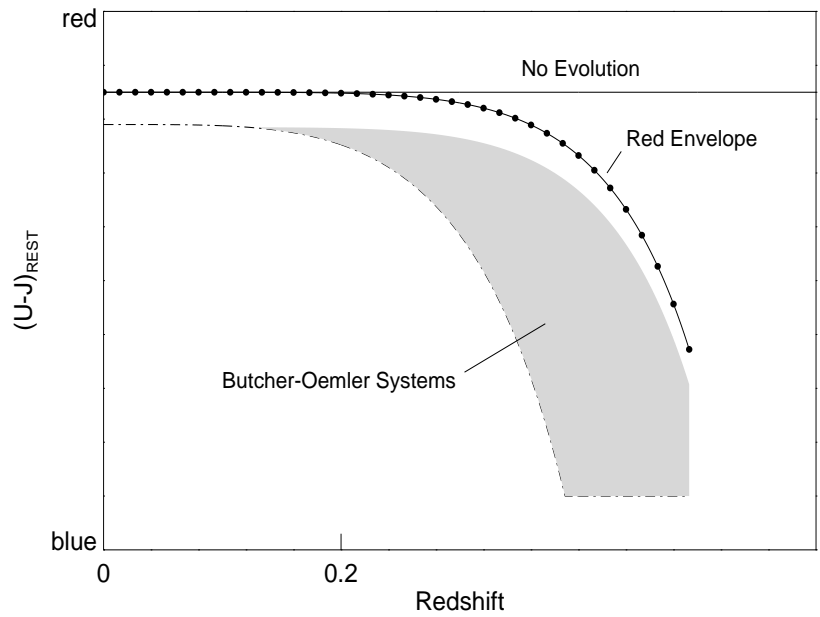,height=3.5in,width=4.6in}}

\caption{ The red envelope test, illustrated for a long-baseline
color measured in the rest frame.  The dashed-dotted line is
the lower boundary of colors for luminous galaxies in rich clusters.
The dotted line is the upper bound, or red envelope.  One measures
the departure of the upper bound from the ``no-evolution'' line and
also its slope at high redshift.  The shaded area shows Butcher-Oemler
systems.    }
\end{figure}

Several groups have now applied the red envelope test to distant
clusters and red field galaxies with $z\,\lsim\,0.9$.  Bruzual
\& Charlot (1993) have summarized 4000 \AA\ Break data from Hamilton
(1985) and Spinrad (1986).  There is a significant decrease with
redshift in the mean and maximum strength of the Break (implying
younger populations), a result confirmed by the independent data set
of Dressler \& Gunn (1990).  Departure of the envelope from the
no-evolution prediction occurs at $z \sim 0.5$.   Using the 
evolutionary models to age-date the Break values, one finds
that 3/4 have formation redshifts above 1, which implies a present age
of $t\,\gsim\,10$ Gyr for H$_0 = 50$ and q$_0 = 0$.  The envelope
itself corresponds to present ages of 13--16 Gyr.  

A systematic approach has been taken by Arag\'on-Salamanca et al.~(1993),
who measured broad-band V,I,K colors in 10 distant rich clusters up to
$z = 0.92$.  They find a clear and systematic trend of envelope color with
redshift, again departing from the no evolution expectation at $z \sim
0.5$.  By $z \sim 0.9$, the envelope is almost undefined; 
galaxies scatter in color, and the characteristic red
peak with a blueward Butcher-Oemler tail has almost vanished.  The
envelope trend corresponds to a formation redshift $\sim 2$, or a
present age of $\sim 13$ Gyr.  The FWHM in the peak at $z = 0.7$,
after removing measurement error, corresponds to a factor of 3
in rest frame age, or a minimum present age of $\sim 10$ Gyr.  

Many detections of red, high $z$ objects are now being reported in
various deep fields.  The red interlopers in nearer clusters suggest
that these systems may not always be dominated by unextincted
starlight.  Recent spectroscopy of high redshift radio galaxies
(Hammer et al.~1993) reveals AGN-like continua in some cases and
demonstrates that caution is needed in trying to age-date distant
field systems from their colors.  

\section{ Conclusion}

Strong evolutionary processes have transformed stellar populations in
galaxies in the relatively recent past ($\lsim$5 Gyr).  The evidence is
unambiguous in the case of distant rich clusters, where disk galaxies are
undergoing starbursts, probably induced by encounters or mergers.  It
is strong for field disks as well.  That this was largely
unanticipated from early studies of nearby galaxies demonstrates the
inadequacy of our older tools for population analysis, particularly
those based on broad-band colors.  

No evidence for a unique galaxy formation epoch has yet emerged.
Definite trends in spectral properties with redshift indicate that the
least active luminous systems formed $\sim$
10--13 Gyr ago (for H$_0 = 50$ and q$_0 = 0$).  Ages from such ``red
envelope'' tests do not necessarily apply to all elliptical galaxies
or to the progenitors of E/S0's in the local, low density
environment.  Judged from both the local and distant samples, galaxy
evolution appears to be faster in denser environments.


\bigskip
\noindent  This research has been supported in part by NASA
Long Term Space Astrophysics grant NAGW-2596.





\vspace{0.15truecm}

\section{Stellar Populations at Large Redshifts: Update (January 1999)}

There has been tremendous progress in the study of stellar populations
at large redshifts in the five years since IAU Symposium 164.  Two key
instruments, the Hubble Space Telescope and the Keck twin 10-m
telescopes, have come into their own, and invaluable faint galaxy
observations have been added by other telescopes, notably the CFHT and
the AAT.  The landmark contribution was the deep, multicolor imaging
obtained by HST of the Hubble Deep Field (Williams et al.\ 1996),
which catalyzed a broad attack on the early evolution of galaxies.
First results of HDF research have been reviewed in two major
meetings (Tanvir et al.\ 1996 and Livio et al.\ 1997).  

The evidence does not support the existence of a unique, well-defined
epoch of galaxy formation, at least insofar as the stellar content of
galaxies is concerned.  To the contrary, the traditional picture (e.g.\
Tinsley 1980) of star formation in galaxies beginning intensely and
synchronously at high redshift and then declining in smooth
exponentials, depending only on galaxy type, is almost certainly
wrong.  Instead, dramatic transformations in star formation histories
(even morphologies) appear to prevail among all types of galaxies and
in all environments, and these can be traced to relatively recent
times ($z \sim 0.2$).  The oldest systems yet age-dated from spectra
are very red radio galaxies at $z \sim 1.5$ with stellar ages $\gsim
3$ Gyr and therefore formation redshifts $z_{\rm F} > 6$ (e.g.\
Peacock et al.\ 1998).  However, the cosmic star formation rate does
not exhibit a pronounced maximum anywhere in the sampled redshift
range.  It rises sharply from low redshifts (Lilly et al.\ 1996) but
then either remains roughly constant for $z \sim 1-4$ (Steidel et al.\
1999) or exhibits a broad peak at an unexpectedly low redshift of $z
\sim 1-2$ (e.g.\ Pei \& Fall 1995, Madau et al.\ 1996), corresponding to a
lookback of 12 Gyr (for H$_0 = 50$ and q$_0 = 0$).

The situation in the field has been thoroughly reviewed by Ellis
(1997).  From HST morphological classifications, the systems
responsible for the excess in the blue counts of faint galaxies are
found to be moderate luminosity, moderate redshift, irregular/peculiar
objects which rapidly disappear at lower redshifts (Driver et al.\
1995, Abraham et al.\ 1996, Brinchmann et al.\ 1998).  Their fate,
either as faded dwarfs or in mergers, has not been conclusively
determined, but the conservative mild- or no-evolution models (\S 2.2)
for the luminosity function seem to be excluded.  Luminous spirals and
ellipticals show less conspicuous evolution to $z \sim 1$.  Truly
quiescent systems with $z_{\rm F} > 3$ would be distinguished by very
red observed colors at $z \sim 1$.  However, the number of such
ultrared objects is small, and their statistics (Kaufmann et al.\
1996, Zepf 1997, Barger et al.\ 1999) suggest that the majority of
field E/S0 galaxies formed more recently than $z \sim 1$ (a lookback
of 10 Gyr).  A significant fraction of morphologically-confirmed field
E/S0's and spiral bulges at $z \sim 0.5-1$ have experienced major star
formation episodes in the preceding 2 Gyr (Abraham et al.\ 1999,
Menanteau et al.\ 1999).  These results contrast with those for denser
cluster environments (see below) but seem consistent with expectations
for hierarchical formation of many ellipticals from disk mergers.  

A remarkably large sample of star-forming systems with $z\, \gsim\, 3$ has
recently been identified, mainly by means of the short-wavelength
``drop\-outs'' produced when the Lyman discontinuity at 912 \AA\ in the
restframe is redshifted into optical-band filters (Steidel et al.\
1996, Lowenthal et al.\ 1997, Giavalisco 1997).  The restframe far-UV
spectra resemble local starforming galaxies, with indications of
interstellar gas outflows but only inconclusive evidence of lower
metal abundances.  Estimated star formation rates are generally $\sim
10-100\, {\rm M}_\odot \ {\rm yr}^{-1}$, much smaller than expected in
the traditional rapid collapse models for spheroid formation.  Because
strong discontinuities in restframe far-UV spectra also occur for
populations with ages up to $\sim 1$ Gyr following the cessation of
star formation, the dropout technique should select such quenched
systems as well.  Their apparent rarity in the spectral samples
implies that continuous star formation persists for long periods,
$>>1$ Gyr, in the dropout systems.  

The dropout systems are intrinsically small, with half-light radii
$\lsim\, 5$ Kpc.  Some could be the cores of present-day spheroids, but
they will likely evolve, probably often by hierarchical accretion, into a
diverse range of descendents.  

Recent studies of distant, rich clusters have confirmed and elaborated
the interpretation discussed in \S2--3 above.  Galaxy evolution is
definitely accelerated, relative to the field, in the denser cluster
environment.  Good HST morphological classifications for cluster
members now enable the identification of a bona fide E/S0 sample for
studies of the ``red envelope,'' which is found to remain well defined
up to $z \sim 0.9$ (Stanford et al.\ 1998).  Mean trends in E/S0
restframe colors with redshift are consistent with passive evolution
and a formation redshift $z_{\rm F} > 1-2$, unlike the behavior in the
field.  Luminous E/S0's show an impressively small scatter in the
photometric-dynamical correlations of the fundamental plane (e.g.\ van
Dokkum \& Franx 1996) and also in restframe color ($\sigma\,
\lsim\, 0.1$ mag in U--V); and the color scatter seems independent of
redshift, cluster richness, or X-ray properties (Ellis et al.\ 1997,
Stanford et al.\ 1998).  The color scatter places a limit on $\delta
t/t$, where t is the age at the observed epoch.  For the quoted
1$\sigma$ value $\delta t/t \sim 0.3$.  This means that the ratio of
maximum to minimum age for the 68\% of the red envelope galaxies
within 1$\sigma$ of the mean is 1.9, assuming that age is entirely
responsible for the spread.  The observations therefore permit a
fairly extended period of E/S0 formation in clusters.  Unless one can
determine the absolute value of $\delta t$ by independent means, the
color scatter does not constrain the age or formation redshift of the
red envelope, though the colors themselves do.  Better constraints on
$z_{\rm F}$ could be placed by pressing the observations farther into
the restframe ultraviolet.  

The high resolution HST data demonstrate that the cluster
density-mor\-phology relation has changed significantly from its
current-epoch character by a redshift of only $z \sim 0.3$ (e.g.
Dressler et al.\ 1997, Couch et al.\ 1998).  There are many more
spirals (Sa-Sdm/Irr) and many fewer S0's (factors of 2-4) than in
nearby rich clusters.  A dominant evolutionary effect in clusters
during the past 5 Gyr has therefore been the transformation of spirals
into S0's.  The Butcher-Oemler population of blue galaxies does consist
mainly of disk systems, but most of these do not have normal
spiral-like spectra (see \S 2.2).  The spectral disturbances
illustrated in Figure 1 do not last long, and Barger et al.\ (1996)
and Couch et al.\ (1998) have shown that the statistics of the burst
and post-burst spectral types imply that a large proportion, roughly
30\%, of the cluster members have suffered major changes in their star
formation rates (bursts or quenching) within the preceding 2 Gyr.
Most of the objects with intense star formation are definitely
undergoing mergers.  Although some of the post-burst objects are
ellipticals, and therefore possibly merger products, many have normal
S0-Sb spiral morphologies.  The data suggest that several processes
besides major mergers, perhaps small accretion events and ram-pressure
stripping, are important in shaping cluster populations.


\begin{thebibliography}{}

\baselineskip=11pt \parskip=0pt


\bibitem{}
Aaronson, M., \& Mould, J. 1980, ApJ, 240, 804

\bibitem{}
Arag\'on-Salamanca, A., Ellis, R.S., Couch, W.J., \& Carter, D.
1993, MNRAS, 262, 764

\bibitem{}
Bender, R., Burstein, D., \& Faber, S.M. 1993, ApJ, 411, 153

\bibitem{}
Bower, R.G., Lucey, J.R., \& Ellis, R.S. 1992, MNRAS, 254, 601

\bibitem{}
Bressan, A., Chiosi, C., \& Fagatto, F. 1994, ApJ, in press

\bibitem{}
Broadhurst, T.J., Ellis, R.S., \& Glazebrook, K. 1992, Nature, 355, 55

\bibitem{}
Bruzual, G., \& Charlot, S. 1993, ApJ, 405, 538

\bibitem{}
Bruzual, G. \& Kron, R.G. 1980, ApJ, 241, 25

\bibitem{}
Butcher, H. \& Oemler, A. 1978, ApJ, 219, 18

\bibitem{}
Chambers, K.C., \& Charlot, S. 1990, ApJ, 384, L1

\bibitem{}
Couch, W.J. \& Sharples, R.M. 1987, MNRAS, 229, 423

\bibitem{}
Couch, W.J., Ellis, R.S., Sharples, R.M., \& Smail, I. 1994, ApJ,
430, 121  

\bibitem{}
Dressler, A. \& Gunn, J.E. 1982, ApJ, 263, 533

\bibitem{}
Dressler, A., \& Gunn, J.E. 1990, in Evolution of the Universe of
Galaxies, ed.~R.G. Kron, (San Francisco: ASP), p. 200

\bibitem{}
Dressler, A., Oemler, A., Butcher, H.R., \& Gunn, J.E.
1994, ApJ, 430, 107

\bibitem{}
Freedman, W.L. 1994, in The Local Group, (CTIO/ESO), in press

\bibitem{}
Fritze-v.~Alvensleben, U., \& Gerhard, O.E. 1994, A\&A, in press

\bibitem{}
Gallagher, J.S., Hunter, D.A., \& Tutukov, A.V. 1984, ApJ,
284, 544

\bibitem{}
Hamilton, D. 1985, ApJ, 297, 371  

\bibitem{}
Hammer, F., Le F\`evre, O., \& Angonin, M.C. 1993, Nature, 362, 324

\bibitem{}
Hubble, E. 1936, The Realm of the Nebulae (New Haven: Yale
Univ. Press)

\bibitem{}
Kennicutt, R.C. 1992, ApJ, 388, 310  

\bibitem{}
Koo, D.C., Gronwall, C., \& Bruzual, G. 1993, ApJ, 415, L21

\bibitem{}
Koo, D.C., \& Kron, R.G. 1992, ARA\&A, 30, 613

\bibitem{}
Larson, R.B., \& Tinsley, B.M. 1978, ApJ, 219, 46

\bibitem{}
Lavery, R.J., Pierce, M.J., \& McClure, R.D. 1992, AJ, 104, 2067

\bibitem{}
Lilly, S.J. 1988, ApJ, 333, 161 

\bibitem{}
Lilly, S.J., Cowie, L.L., \& Gardner, J.P. 1991, ApJ, 369, 79

\bibitem{}
O'Connell, R.W. 1980, ApJ, 236, 430

\bibitem{}
O'Connell, R.W. 1988, in Towards Understanding Galaxies at High
Redshift, eds.~A. Renzini \& R.G. Kron (Dordrecht: Kluwer), p. 177

\bibitem{} 
O'Connell, R.W. 1994, in Nuclei of Normal Galaxies: Lessons
from the Galactic Center, ed. R. Genzel (Dordrecht: Kluwer), in press 

\bibitem{}
Oemler, A. 1992, in Clusters \& Superclusters of Galaxies,
ed.~A.C. Fabian (Dordrecht: Kluwer), p. 29

\bibitem{}
Pickles, A.J., \& van der Kruit, P.C., A\&AS, 84, 421

\bibitem{}
Rose, J.A. 1994, AJ, 107, 206

\bibitem{}
Sandage, A. \& Visvanathan, N. 1978, ApJ, 223, 707

\bibitem{}
Schweizer, F., \& Seitzer, P. 1992, AJ, 104, 1039

\bibitem{}
Silva, D.R. 1991, PhD Thesis, Univ. of Michigan

\bibitem{}
Spinrad, H. 1986, PASP, 98, 269

\bibitem{}
Tinsley, B.M. 1980, ApJ, 241, 41

\bibitem{}
Tyson, J.A. 1988, AJ, 96, 1

\bibitem{}
Tyson, J.A., \& Jarvis, J.F. 1979, ApJ, 230, L153

\bibitem{}
Worthey, G. 1994, ApJS, in press

\bibitem{}
Wu, C.-C., Faber, S.M., Gallagher, J.S., Peck, M., \& Tinsley, B.M.
1980, ApJ, 237, 290


\end{thebibliography}

\begin{thebibliography}{}

\centerline{\bf (Updates for 1995-98)}

\baselineskip=11pt \parskip=0pt


\bibitem{} 
Abraham, R.G., Ellis, R.S., Fabian, A.C., Tanvir, N.R., and Glazebrook, K.
1999, {MNRAS}, in press

\bibitem{} 
Abraham, R.G., Tanvir, N.R., Santiago, B., Ellis, R.S., Glazebrook,
K., and van den Bergh, S. 1996, {MNRAS}, {279}, L47

\bibitem{} Barger, A.J., Arag\'on-Salamanca, A., Ellis, R.S., Couch,
W.J., Smail, I., and Sharples, R.M. 1996, {MNRAS}, {279}, 1

\bibitem{} 
Barger, A.J., Cowie, L.L., Trentham, M., Fulton, E., Hu, E.M.,
Songaila, A., and Hall, D. 1999, {AJ}, in press

\bibitem{} 
Brinchmann, J., Abraham, R., Schade, D, Tresse, L, Ellis, R.S., 
et al.\ 1998, {ApJ}, {499}, 112


\bibitem{}
Couch, W.J., Barger, A.J., Smail, I., Ellis, R.S., and Sharples,
R.M. 1998, {ApJ}, {497}, 188

\bibitem{}
Dressler, A., Oemler, A., Couch, W.J., Smail, I., Ellis, R.S., et al.
1997, {ApJ}, {490}, 577

\bibitem{} 
Driver, S.P. , Windhorst, R.A., Ostrander, E.J., Keel, W.C.,
Griffiths, R.E., and Ratnatunga, K.U. 1995, {ApJ}, {449}, L23

\bibitem{} 
Ellis, R.S. 1997, {Ann. Rev. Astr. Ap.}, {35}, 389

\bibitem{ }
Ellis, R.S., Smail, I., Dressler, A., Couch, W.J., Oemler, A., et al.
1997, {ApJ}, {483}, 582

\bibitem{} Giavalisco, M. 1997, in { The Hubble Deep Field}, eds. M.
Livio, S.M. Fall, and P. Madau, (Cambridge: Cambridge University
Press), p. 121

\bibitem{} 
Kauffmann, G., Charlot, S., and White, S.D.M. 1996, {MNRAS},
{283}, L117

\bibitem{} 
Lilly, S.J., Le F\`evre, Hammer, F., and Crampton, D. 1996,
{ApJ}, {460}, L1

\bibitem{} 
Livio, M., Fall, S. M., and Madau, P. (eds) 1997, {The Hubble
Deep Field} (Cambridge: Cambridge University Press)

\bibitem{ }
Lowenthal, J.D., Koo, D.C., Guzman, R., Gallego, J., Phillips, A.C., et al.
1997, {ApJ}, {481}, 673

\bibitem{} 
Madau, P., Ferguson, H.C., Dickinson, M.E., Giavalisco, M.,
Steidel, C.C., and Fruchter, A. 1996, {MNRAS}, {283}, 1388

\bibitem{} 
Menanteau, F., Ellis, R.S., Abraham, R.G., Barger, A.J., and
Cowie, L.L. 1999 {MNRAS}, in press

\bibitem{} 
Peacock, J.A., Jimenez, R., Dunlop, J.S., Waddington, I., Spinrad, H.,
et al. 1998, {MNRAS}, {296}, 1089

\bibitem{} 
Pei, Y.C., and Fall, S.M. 1995, {ApJ}, {454}, 69

\bibitem{ } 
Stanford, S.A., Eisenhardt, P.R., and Dickinson, M. 1998, {ApJ},
{492}, 461

\bibitem{}
Steidel, C.C., Adelberger, K.L., Giavalisco, M., Dickinson, M.,
and Pettini, M. 1999, {ApJ}, in press

\bibitem{}
Steidel, C.C., Giavalisco, M., Pettini, M., Dickinson, M., and
Adelberger, K.L. 1996, {ApJ}, {462}, L17

\bibitem{} 
Tanvir, N.R., Arag\'on-Salamanca, A. and Wall, J.V., (eds) 1997, 
{HST and the High Redshift Universe}, (Singapore: World Scientific Press)

\bibitem{ }
van Dokkum, P.G., and Franx, M. 1996, {MNRAS}, {281}, 985

\bibitem{} 
Williams, R.E., Blacker, B., Dickinson, M., Van Dyke Dixon, W.,
Ferguson, H. C., et al. 1996, {AJ}, {112}, 1335

\bibitem{} 
Zepf, S.E. 1997, {Nature}, 390, 377


\end{thebibliography}
\end{document}